\documentclass{revtex4-2} 
\usepackage{url,hyperref,lineno,microtype,subcaption}
\usepackage[onehalfspacing]{setspace}
\usepackage{amssymb}
\usepackage{amsfonts}
\usepackage{amsmath}
\usepackage{graphicx}
\usepackage{bm}

\DeclareMathOperator{\im}{Im}
\DeclareMathOperator{\re}{Re}

\begin{document}

\author{Vadim Roytershteyn}
\affiliation{Space Science Institute, Boulder, CO, USA}
\author{Gian Luca Delzanno} 
\author{Justin C. Holmes}
\affiliation{T-5 Applied Mathematics and Plasma Physics Group, Los Alamos National Laboratory, Los Alamos, NM, USA }

\title[Oblique Instability of Quasi-Parallel Whistler Waves]{Oblique Instability of Quasi-Parallel Whistler Waves in the Presence of Cold and Warm Electron Populations} 

\begin{abstract}

Whistler waves propagating nearly parallel to the ambient magnetic field experience a nonlinear instability due to transverse currents when the background plasma has a population of sufficiently low energy electrons. { Intriguingly, this nonlinear process may generate oblique electrostatic  waves, including whistlers near the resonance cone with properties resembling oblique chorus waves in the Earth's magnetosphere. Focusing on the generation of oblique whistlers, earlier analysis of the instability is extended here to the case where low-energy background plasma consists of both a "cold" population with energy of a few eV and a "warm" electron component with energy of the order of 100 eV.} This is motivated by spacecraft observations in the Earth's magnetosphere where oblique chorus waves were shown to interact resonantly with the warm electrons. The main new results are: i) the instability producing oblique electrostatic waves is sensitive to the shape of the electron distribution at low energies. { In the whistler range of frequencies, two distinct peaks in the growth rate are typically} present for the model considered: a peak associated with the warm electron population at relatively low wavenumbers and a  peak associated with the cold electron population at relatively high wavenumbers; ii) overall, the instability producing oblique whistler waves near the resonance cone persists (with a reduced growth rate) even in the cases where the temperature of the cold population is relatively high, including cases where cold population is absent and only the warm  population is included; iii) particle-in-cell simulations show that the instability leads to heating of the background plasma and formation of characteristic plateau and beam features in the parallel electron distribution function in the range of energies resonant with the instability. The plateau/beam features have been previously detected in spacecraft observations of oblique chorus waves. However, they have been attributed to external sources and have been proposed to be the mechanism generating oblique chorus. In the present scenario, the causality link is reversed and the instability generating oblique whistler waves  is shown to be a possible mechanism for formation of the plateau and beam features.

\end{abstract}

\maketitle

\section{Introduction}

Whistler waves play a critical role in the dynamics of the Earth's magnetosphere since wave-particle interactions with whistlers play a dual role of both accelerating and scattering of energetic electrons. The corresponding diffusion coefficients depend on the global distribution of spectral power and polarization of the waves, making understanding of these properties  essential for successful modeling of ring current or radiation belt dynamics~\citep{Thorne_2010}. As a consequence, there exists a significant interest in understanding of the basic physical processes affecting evolution of the magnetospheric waves in general and of whistler waves in particular. 

Statistical studies show that whistler waves in the Earth's magnetosphere appear either as ``chorus'', i.e. discrete emissions appearing typically in two distinct frequency bands in a region outside the plasmasphere~\citep{burtis69,tsurutani74,li2013}, or ``hiss'', i.e.  broadband emissions found predominantly in the plasmasphere ~\citep{dunckel69,russell69,thorne73,hartley18} and in plasmaspheric plumes~\citep{Chan1976}.  The chorus waves are associated with local energization of energetic particles~\citep{summers1998,meredith02,meredith03}, as well as electron precipitation in the form of diffuse~\citep{Swift_1981,Ni14} and pulsating aurora~\citep{nishimura10,kasahara2018}, or microbursts~\citep{oliven68,breneman17}. Chorus waves observed in the Earth's magnetosphere can reach very large amplitudes, e.g.~\citep{Cattell2008,Wilson2011,Tyler2019a,Tyler2019b}.  For small magnetic latitudes, lower-band chorus waves are predominantly field-aligned~\citep{hayakawa1984,haque2010chorus,li2011chorus,agapitov2012corr,li2013,teng2019}, which is consistent with their excitation via a cyclotron instability of anisotropic electrons with energies in the range of several to tens of keV~\citep{kennel66,gary96,gary11,fu2014}.  In addition to the field-aligned chorus waves, waves near the resonance cone angle are often present~\citep{santolik09,agapitov2012corr,li2013,li_2016,teng2019}. The oblique waves play an important role in scattering of energetic electrons~\citep{mourenas2014,Artemyev2015}. In general, the whistler waves become more oblique as they propagate to high latitudes. They may further experience significant radial propagation, which complicates identification of their sources~\citep{chen2013}. Nevertheless, it is likely that the oblique chorus at small magnetic latitudes is generated by a mechanism distinct from that producing field-aligned chorus~\citep{Artemyev2015}. Analyses of specific events often find plateau/beam-like features in the parallel electron energy distribution with typical energy from $\sim 100$ eV~\citep{li_2016} to several keV~\citep{agapitov2015}.  This led to suggestions that oblique modes are generated by a  linear instability associated either with temperature anisotropy of $\sim$ keV electrons plus a plateau/beam distribution at lower energies or with a beam-plasma instability driven by low-energy ($\lesssim 1$ keV) beams~\citep{mourenas2015,Artemyev2016,li_2016}. An alternative mechanism was proposed by~\cite{Fu2017}, { who} suggested that oblique chorus waves are generated by wave-wave nonlinear coupling from the interaction of quasi-field-aligned chorus waves in the lower band and mildly oblique upper-band chorus waves.

Since chorus waves are typically observed in the regions adjacent to the plasmasphere, the density of cold plasma populations in these regions is often significant ({ here "cold" is loosely defined to mean energies below approximately 100eV.)} In fact, in cases where the total plasma density can be measured by spacecraft (e.g. from  the frequency of the upper-hybrid line), it is often found that cold plasma populations dominate the total density. The focus of this paper is on a recently identified nonlinear instability experienced by whistler waves in the presence of cold plasma populations~\citep{roytershteyn2021}. Specifically, kinetic simulations and theory were used to show that in the presence of cold electron populations, quasi-field-aligned whistler waves of sufficiently large amplitude experience a parametric nonlinear instability. This process leads to damping of quasi-field-aligned waves,  heating of the cold plasma, and generation of oblique electrostatic modes, including oblique whistlers waves near the resonance cone. The latter in particular are of significant interest since the instability in question may provide a new mechanism of generating oblique chorus that is complementary to the earlier proposed models. In this paper, we extend the analysis of the instability to the case of a multi-component distribution of low-energy electrons{, focusing specifically on the generation of the oblique whistlers.  Inclusion of such additional electron populations} is motivated by the need to make a closer contact with observations that often show presence of ``warm'' electron populations  with energies in the range of $\sim 100$ eV, which can also strongly interact with the oblique chorus waves{ ~\citep[e.g.][]{agapitov2015,li_2016,li_origin_2019}}. Using theoretical analysis and kinetic particle-in-cell (PIC) simulations, we show that the previously identified instability producing oblique whistler waves is robust in the sense that is remains operational in the presence of such warm electron populations. Further, it may produce observed features of the electron distributions, such as formation of a plateau or a weak beam in the resonant range of energies. 

{ Properties of cold plasma populations, especially of the electrons, are  hard to measure because their typical energies are comparable to the spacecraft potential and/or the energy of the photo-electrons emitted by the spacecraft~\citep[see e.g., a review in][]{delzanno2021}. Photo-electrons, in particular, strongly confound the measurements of ambient cold electrons for spacecraft in sunlight. As a result, the energy distribution of the cold  populations remains poorly constrained by observations even if the total plasma density could sometimes be estimated from wave measurements.  In the present work, we treat the temperature of the cold populations as an unknown and vary it parametrically.}

In what follows, we use the following notations: the background magnetic field ${\bm B}_0$ is along z in a local Cartesian $(x,y,z)$ system. Perpendicular and parallel directions are defined with respect to $\bm B_0$. The plasma frequency for species $s$ is defined by $\omega^2_{p,s}={4 \pi e^2 n_s}/{m_s}$, with $n_s$ the density of species $s$, $e$ is the absolute value of the electron charge, and $m_s$ is the species mass.  Index $s=eC,\,eW,\,eH,\,i$ labels cold (eC), warm (eW) and hot (eH) electrons and ions (i). { Here "hot" refers to an anisotropic population with energies in the range of several keV and above.} The electron plasma frequency computed with the total electron density $n_0=n_{0eC}+n_{0eW}+n_{0eH}$ is denoted by $\omega_{pe}$ and the total electron inertial length is $d_e=c/\omega_{pe}$ with $c$ the speed of light. The cyclotron frequency is $\Omega_{c,s}=eB_0/m_s c$. The species thermal velocity is $v_{t,s}=\sqrt{{T_s}/{m_s}}$, where $T_s$ is the temperature (which could be anisotropic). The inertial length is $d_s=c/\omega_{p,s}$. We will interchangeably use the terms ``parent'', ``driver'', or ``primary'' wave to refer to quasi field-aligned (whistler) wave that is the source of energy for the nonlinear interactions considered.

\section{Linear Theory\label{sec:linear}}
The specific nonlinear mechanisms that are the focus of this investigation originate from relative drifts between cold electrons and ions transverse to the background magnetic field. In principle, it is well known that transverse drifts, when present,  may produce instabilities that play an important role in many settings in laboratory and space plasmas (see e.g. a review of instabilities in the shock foot by~\cite{Muschietti2017}). In the cases discussed here, the drifts are induced by the oscillating electric field of the parent whistler wave. The key observations are: i) whistlers are often present in the region where density is dominated by cold plasma populations and ii) for large, but realistic amplitude of whistlers (i.e. comparable to the observed { values}), the drifts can be significant in relation to the thermal speed of the cold electron component. 

The kinetic theory to study the instabilities of quasi-{ parallel} whistler waves caused by relative transverse drift was presented by~\cite{roytershteyn2021}. The key features of the theory are: i) the underlying equilibrium corresponds to a time-dependent circularly-polarized flow of cold plasma populations induced by the primary whistler wave; ii) The dispersion relation for a given wavevector $\bm k$ couples several frequencies, separated by harmonics of the driver frequency $\omega_0$; iii) For typical parameters of magnetospheric plasmas, the dominant instabilities are  either quasi-perpendicular, short-wavelength modes related to Electron Cyclotron Drift Instability~\citep{Forslund1970,Forslund1972}, or oblique short-wavelength whistler modes near the resonance cone. 

Specifically, the electrostatic dispersion relation, generalized here to account for multiple electron species, reads:
\begin{eqnarray}
&&  \left \{ k^2 +  \sum_{s=eC,eW}
\frac{\omega_{pe,s}^2}{v_{t,s}^2} \sum_{n=-\infty}^{+\infty} \left[ 1+\frac{\omega}{\sqrt{2} v_{t,s} k_z} Z\left( \xi_{s,n} \right)\right]  \Gamma_n \left( \lambda_s \right)   \right \}\tilde{\phi}({\bm k},\omega) 
  \nonumber \\ 
&&=
\frac{\omega_{pi}^2}{2 v_{ti}^2} \sum_{p,m=-\infty}^{+\infty} J_p\left( a \right) J_m\left( a \right) Z^\prime\left( \frac{\omega+p\omega_0}{\sqrt{2} k v_{ti}}\right) \tilde{\phi}({\bm k},\omega+(p-m)\omega_0). \nonumber \\
\label{eq:disp}
\end{eqnarray}
Expression \eqref{eq:disp} is written in the electron co-moving (rotating) frame of reference. Here $\omega$ is the frequency, $k=|\bm k|$ is the magnitude of the wavevector $\bm k = \{0,k_y,k_z\}$, $a= {k_y V_0}/{\omega_0}$,  $\Gamma_n (\lambda) = e^{-\lambda}  I_n \left( \lambda\right)$, $\lambda_s = k_y^2 v_{t,s}^2/\Omega_{ce}^2$, $\xi_{s,n} = (\omega-n\Omega_{ce})/(\sqrt{2} v_{t,s} k_z)$ ,  $Z$ is the plasma dispersion function,  $J_m$ ($I_n$) is the (modified) Bessel function of first kind, and $\tilde{\phi} (\bm k, \omega)$ is the Fourier component of the electrostatic potential in that frame,  while $V_0$ and $\omega_0$ are the amplitude and frequency of the equilibrium electron flow induced by a primary whistler wave. Eq.~\eqref{eq:disp} is derived under assumption that the spatial variation of the driving quasi-parallel whistler mode can be neglected. This is rigorously justified under condition $k_z \gg k_{z0}$, where $k_{z0}$ is characteristic wavenumber of the driving mode. { We note that in deriving Eq.~\ref{eq:disp}, we assumed that the cold and warm populations are isotropic. Accounting for anisotropy is straighforward: corresponding expressions describing electron response (in non-rotating frame) are given in many textbooks, e.g.~\cite{Stix1992}.}

As is emphasized above, expression~\eqref{eq:disp} couples the response of the system at a given ${\bm k}$ and $\omega$ with that at the same ${\bm k}$ but at at different values of $\omega$, Doppler-shifted by the frequency of the driver $\omega_0$. Thus, relation \eqref{eq:disp} constitutes a system of equations for $\omega$ and sidebands $\omega\pm p \omega_0$, where $p$ is integer. However, when $|\omega+p\omega_0|\gg k v_{ti}$ the ion contribution (terms on the right-hand side of equation \eqref{eq:disp}) can be neglected. In this  case, system \eqref{eq:disp} describes the usual electron waves in a hot plasma, expressed however in the rotating frame. Instabilities may appear near intersections of a specific hot-plasma electron wave with the Doppler-shifted ion response, i.e. when { $|\omega+p\omega_0| \lesssim k v_{ti}$ }. Finally, the transformation between Fourier components of any quantity $A$ in the co-moving frame $\tilde A$ and those in the stationary (ion) frame $\hat A$ is given by 
$ \tilde A (\omega, k) = \sum_m J_m(a) \hat A(\omega + m \omega_0 , k )$~\citep{Kaw1973}. Since observations are made in the stationary frame, this implies that a mode with a single frequency $\omega$ in the co-moving frame may be observed as a series of sidebands.

A variety of useful approximations to Eq.~\eqref{eq:disp} could be readily obtained. First, we ignore coupling to the side-bands and focus on perturbations in the vicinity of the resonance cone $\omega \approx \Omega_{ce} \cos \theta$ to obtain a simplified dispersion relation (c.f. Eq. 7 of \cite{roytershteyn2021}):
\begin{equation} 
k^2 +  \sum_{s=eC,eW}
\frac{\omega_{pe,s}^2}{v_{t,s}^2} \sum_{n=-\infty}^{+\infty} \left[ 1+\frac{\omega}{\sqrt{2} v_{t,s} k_z} Z\left( \xi_{e,n} \right)\right]  \Gamma_n \left( \lambda \right)  = \frac{\omega_{pi}^2}{2 v_{ti}^2} |J_1(a) |^2 Z^\prime\left( \frac{\omega-\omega_0}{\sqrt{2} k v_{ti}}\right)  \\
\label{eq:disp_simple}
\end{equation}
A justification of the assumptions leading to Eq.~\ref{eq:disp_simple} was discussed by~\cite{roytershteyn2021}. 

A further insight into the nature of the instability generating oblique whistler modes can be obtained by { by considering a configuration where the warm population is absent and} using cold plasma approximation for the left-hand side,  $|\Omega_{ce}/(k_y v_{te})| \gg 1$ and $|\Omega_{ce}/(k_z v_{te})| \gg 1$.  The system \eqref{eq:disp} then reduces to 
\begin{equation}
    k_y^2 \left(1-\frac{\omega_{pe}^2}{\omega^2-\Omega_{ce}^2} \right)+k_z^2\left(1-\frac{\omega_{pe}^2}{\omega^2}\right)-\frac{\omega_{pi}^2}{2 v_{ti}^2}J_{-1}^2\left(a\right)Z^\prime\left(\zeta\right)=0,
    \label{eq:disp_an1}
\end{equation}
with $\zeta=(\omega-\omega_0)/(\sqrt{2} k v_{ti})$, and $a = k_y |V_0|/\omega_0$.
Expanding the Z function in the small argument limit, $Z^\prime\sim -2 i\sqrt{\pi}\zeta \exp(-\zeta^2)-2$, and assuming  $\gamma/\omega_r \ll 1$, where $\gamma = \im \omega$ and $\omega_r = \re \omega$,  yields an analytic expression for the instability growth rate:  
\begin{equation}
   \frac{\gamma}{\omega_r} = \frac{\sqrt{\pi}}{2}\frac{m_e}{m_i} \frac{J_{-1}^2(a)}{v_{ti}^2}  \frac{\omega_0-\omega_r}{\sqrt{2} k v_{ti}}  \left [  \frac{ k_y^2 \omega_r^2}{(\omega_r^2-\Omega_{ce}^2)^2} +\frac{k_z^2}{\omega_r^2} \right ]^{-1}
       \label{eq:disp_an2}
\end{equation}
where $\omega_r$ is the solution of 
\begin{equation}
  k_y^2 \left(1-\frac{\omega_{pe}^2}{\omega_r^2-\Omega_{ce}^2} \right)+k_z^2\left(1-\frac{\omega_{pe}^2}{\omega_r^2}\right)+\frac{\omega_{pi}^2}{ v_{ti}^2}J_{-1}^2\left(a\right)=0. \label{eq:wr_approx}
\end{equation}
The nature of the instability now becomes evident. When there is no flow, the terms involving $J_{-1}$ are zero and there is no coupling between electrons and ions. In that case there is no instability and  Eq. (\ref{eq:wr_approx}) describes whistler waves near the resonance cone angle, $\cos \theta_R \approx \omega_r/\Omega_{ce}$. The presence of the flow leads to coupling with the ions and to a drift instability when $\omega_r<\omega_0$ which leads to the generation of the oblique whistler waves near the resonance cone. 

\begin{figure}[h!]
\centering
\includegraphics[width=30pc]{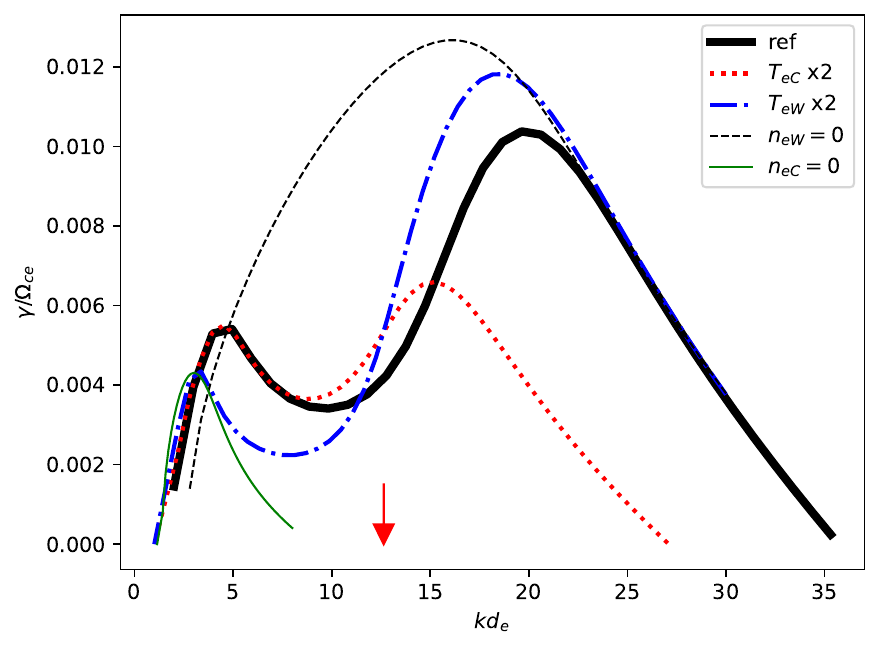}
\caption{Linear theory: effect of the warm electron population. The figure shows dispersion relation $\gamma(k)$ obtained by solving Eq.~\ref{eq:disp_simple} for angle $\theta$  corresponding to the resonance cone $\cos \theta = \omega_0/\Omega_{ce}$, for reference case with the parameters indicated in the text, two cases with the temperatures of the cold and warm population increased by a factor of 2 compared to the reference case, and for the cases where the warm or cold populations are absent. The wavenumber corresponding to condition $|\xi_{eW,0}|= 1$ is indicated by the red arrow at $k d_e \approx 12.6$.  }
\label{fig:warm_dispersion}
\end{figure}

In order to understand the influence of the low-energy part of the electron energy spectrum on the instability, we turn to the form of the dispersion relation given by Eq.~\ref{eq:disp_simple}, which retains the resonance terms. An example of its solution is presented in Fig.~\ref{fig:warm_dispersion}. Here, we assumed parameters approximately corresponding to simulation case D discussed in Section~\ref{sec:simulations} below: $\omega_{pe}/\Omega_{ce}=4$, $n_{0eW}/n_0=0.2$, $n_{0eC}/n_0=0.75$, $T_{eC} = T_i = 5$eV, $T_{eW} = 100$eV, $\omega_0 = 0.3 \Omega_{ce}$,  $V_0 \approx 0.8 v_{teC}$, and $m_i/m_e=1836$. We observe that in the presence of warm and cold populations, there are typically two distinct peaks in the growth rate, one at relatively low values of $k$ and the other at relatively high values of $k$. 
{ The nature of these two peaks could be elucidated by varying the parameters of the background plasma. For example, increasing the temperature of the cold populations by a factor of two strongly affects the peak at higher values of $k$, but leaves the peak at low values of $k$ virtually unchanged (see curve labeled $T_{eC} \times 2$ in Fig.~\ref{fig:warm_dispersion}. In contrast, increasing the temperature of the warm population affects both peaks, significantly reducing the growth rate at the low values of $k$ (curve labeled $T_{eW} \times 2$). Examination of the case where the warm population is absent (curve labeled $n_{eW}=0$) suggests that the peaks are formed due to suppression of the growth rate in the range of wavenumbers where the warm population is resonant with the mode $|\xi_{eW,0}|=1$.
}

Overall, the results of the linear analysis illustrate strong dependence of the instability properties on the energy spectrum of the electrons in the range below a few hundred eV. Further, it may be anticipated that development of the instability may influence distribution functions in this part of the energy spectrum, leading for example to formation of plateau and/or beam features in the distribution due to resonances. This point is illustrated further in the following section using PIC simulations. 

We also note that the oblique instability near the whistler resonance cone persists (for a fixed driver amplitude) for relatively large values of electron temperature. For the parameters considered in the example above, that included cases where the cold population was absent and only the warm population with $T_{eW} = 100$eV  was included, as shown in Fig.~\ref{fig:warm_dispersion}. While the corresponding growth rate becomes lower with increasing temperature of the background populations, this behavior is in contrast to the strong temperature dependence of quasi-perpendicular ECDI-type modes. The latter generally appear when the magnitude of the transverse drift $V_0$ is comparable to the thermal speed of the low-energy populations $v_{te}$. The oblique instability of the whistler type clearly exist in the regime $V_0 < v_{te}$. This observation suggests that the oblique chorus generated by the mechanism presented here could often accompany large-amplitude quasi-field aligned chorus waves in the Earth's magnetosphere.

\section{Evolution of the oblique instabilities and energization of the background plasma: PIC simulations \label{sec:simulations}}

\subsection{Simulation model}

In the simulation model presented here, the field-aligned whistler waves are excited by cyclotron (temperature anisotropy) instability of an electron population with the energy in the range of $\sim 10$ keV. The initial conditions correspond to plasma with protons and up to three species of electrons with different energies. All populations are initially characterized by (bi-) Maxweillian distributions with density $n_{0s}$, parallel (along the background magnetic field) temperature $T_{s||}$ and perpendicular temperature $T_{s\perp}$. The hot ($\sim$ keV) electrons are characterized by the anisotropy $A={T_{eH\perp}}{/T_{eH||}}-1$. The cold ($\sim$ eV) and warm ($\sim$ 100 eV) electrons and the ions are isotropic and hence can be identified by a single temperature, $T_{eC}$, $T_{eW}$ and $T_i$. Simulations are performed with the Particle-In-Cell (PIC) code VPIC~\citep{Bowers2008} using the parameters summarized in Table \ref{tab:params}. The simulations are performed in a two-dimensional Cartesian domain $(y,z)$ of dimension $L_y\times L_z$, discretized by $N_y \times N_z$ cells and use a uniform time step $\Delta t$.   In all cases $\omega_{pe}/\Omega_{ce}=4$ and $n_i/n_0=1$. 

We mainly vary the temperature of the cold electrons and ions relative to the 100 eV warm electrons to investigate the influence of cold plasma temperature on the growth of oblique whistlers and ECDI. The simulation domain size was selected to accommodate the growth of waves expected from the linear theory { described} in the previous section. Some additional runs were performed to investigate a fully 3D system, and special cases with no warm electrons present.

\subsection{Simulation results}

All of the simulations summarized here behave qualitatively in the same way. 
{ The initial configuration is unstable to a cyclotron instability due to the presence of an anisotropic hot electron population. This instability quickly creates the primary, predominantly field aligned, whistler waves. After the primary whistler waves reach saturation, electrostatic fluctuations with finite $k_\perp$ appear, which are driven by the nonlinear processes discussed in Section~\ref{sec:linear}. For the parameters considered, the oblique fluctuations are related to electron Bernstein modes and  oblique electrostatic whistler modes near the resonance cone. These fluctuations lead to damping of the primary whistler modes and heating of the cold and warm electrons. For most of the simulations considered in this study, the most intense transfer of energy from the primary whistler to the low-energy electron populations is associated with the oblique electrostatic whistlers. The final state of the system is characterized by the presence of the oblique whistler modes. An important observational property of these waves is that they have nearly the same frequency as the primary, field-aligned whistlers, but higher wavenumbers and wave-normal angles near the resonance cone.}

\begin{figure}[h!]
\centering
\includegraphics[width=20pc]{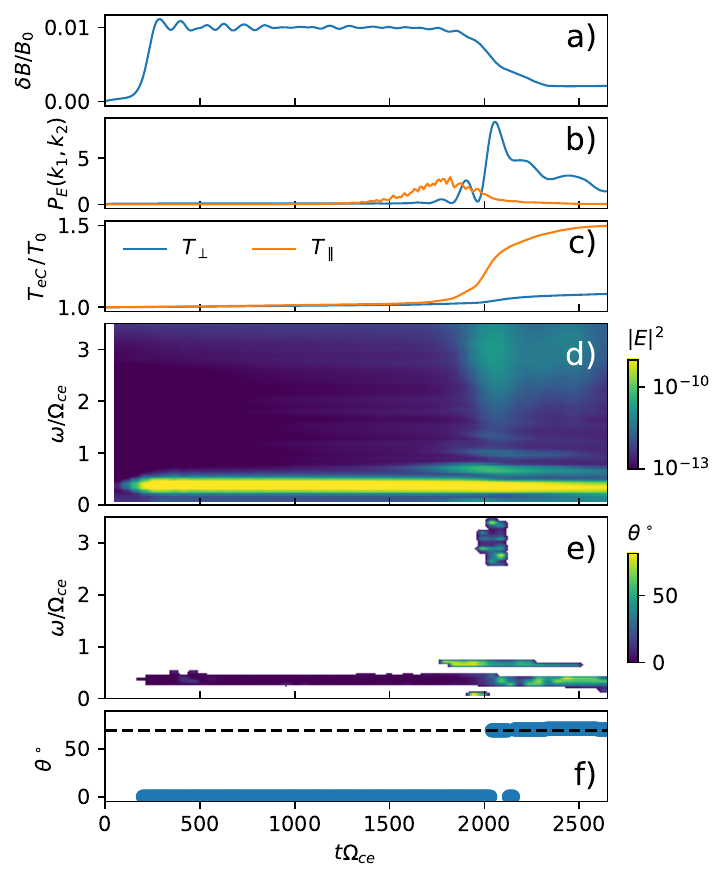}
\caption{Time evolution in simulation D: (a) amplitude of the transverse magnetic field perturbation $\delta B$ relative to the background magnetic field $B_0$;  (b) energy density of electric field fluctuations { $|\bm E|^2$ in a narrow range of wavenumbers corresponding to the highly oblique whistler waves (blue) and the electron Bernstein waves (orange, multiplied by 10 for clarity)};
(c) temperature of cold electrons { normalized to the initial cold temperature $T_0=5$eV}; (d) spectrogram of electric field fluctuations $|E|^2$; (e) wave normal angle spectrum obtained by the SVD method; (f) wave normal angle corresponding to the peak of the spectrum in panel (d). The dashed line corresponds to the resonance angle $\theta_R=69.5^\circ$. { See text for details}} 
\label{fig:time_trace}
\end{figure}

{
In order to illustrate this typical behavior, we focus on simulation D (see Table~\ref{tab:params}).  Fig.~\ref{fig:time_trace} summarizes time-evolution of several relevant diagnostics. Panel a) shows evolution of the average magnitude of the magnetic field fluctuations. The energy density $ P_E $ of the electric field fluctuations with finite $k_\perp$  is shown in Panel b). To differentiate between electron Bernstein and oblique electrostatic whistler waves, $ P_E $ is computed in two narrow ranges of perpendicular wavenumbers $ P_E (k_1,k_2) = \sum_{k_\perp=k_1}^{k_2} |{\bm E}(\bm k)|^2$, where $ |{\bm E}(\bm k)|^2= |E_x(\bm k) |^2 + |E_y(\bm k) |^2 + |E_z(\bm k) |^2$. Panel c) shows parallel and perpendicular cold electron temperatures. The spectrogram of the electric field fluctuations is shown in panel  d). This is computed by performing Fast Fourier Transform (FFT) in a sliding time window of width $100\Omega_{ce}^{-1}$, averaged over y and z. Panel e) shows the wave normal angle computed using singular value decomposition (SVD) of the electromagnetic wave correlation matrix~\citep{Santolik2003} in the same sliding windows. The results are only shown in the areas where the spectrum density $S_E(\omega, t)$ in panel d) is larger than a chosen threshold. To complement this estimate, Fig. \ref{fig:time_trace} (f) shows the wave normal angle for the peak of the wavenumber spectrum at a given time.

As is seen in panel a) of Fig.~\ref{fig:time_trace}, the primary whistler waves in simulation D grow on a timescale of a few hundred $\Omega_{ce}^{-1}$. The waves are characterized by frequency $\omega \approx 0.35\Omega_{ce}$ (see panel d) and are field-aligned (see panels e and f). Electrostatic fluctuations with finite $k_\perp$ appear in the electric field at $t\Omega_{ce}\sim1500$. The orange line in panel b) of Fig.~\ref{fig:time_trace} corresponds to quasi-perpendicular electron Bernstein-like waves ($P_E$ is computed in the range $100 \leq k_\perp d_e \leq 120$) and is multiplied by 10 for clarity, while the blue line corresponds to the highly oblique whistler waves near the resonance cone and is obtained by taking $2 \pi d_e/L_\perp \leq k_\perp d_e \leq 50$ while computing $P_E$. Note that when the electrostatic fluctuations appear, the amplitude of the primary whistler waves begins to decrease (panel a). The most intense damping coincides with excitation of the highly oblique whistlers. The final amplitude of the magnetic field fluctuations associated with the primary whistler waves is $20\%$ of the initial saturated amplitude. Moreover, as the primary whistler waves damp, cold electrons heat, as seen in  Panel c) of Fig. \ref{fig:time_trace}. After approximately $t\Omega_{ce}=2000$, the spectrum is  dominated by the highly oblique waves at the same frequency of the initial parent wave (see panels e and f). Harmonics of the initial wave frequency $\omega_0$ are clearly evident in panel (d), corresponding to both the highly oblique whistlers and the electron Bernstein waves. In particular, the transition at around $t\Omega_{ce}=2000$, when the amplitude of the secondary modes is the highest (see panel b), is characterized by a rich spectral signature. 
}

\begin{figure}[h!]
\centering
\includegraphics[width=20pc]{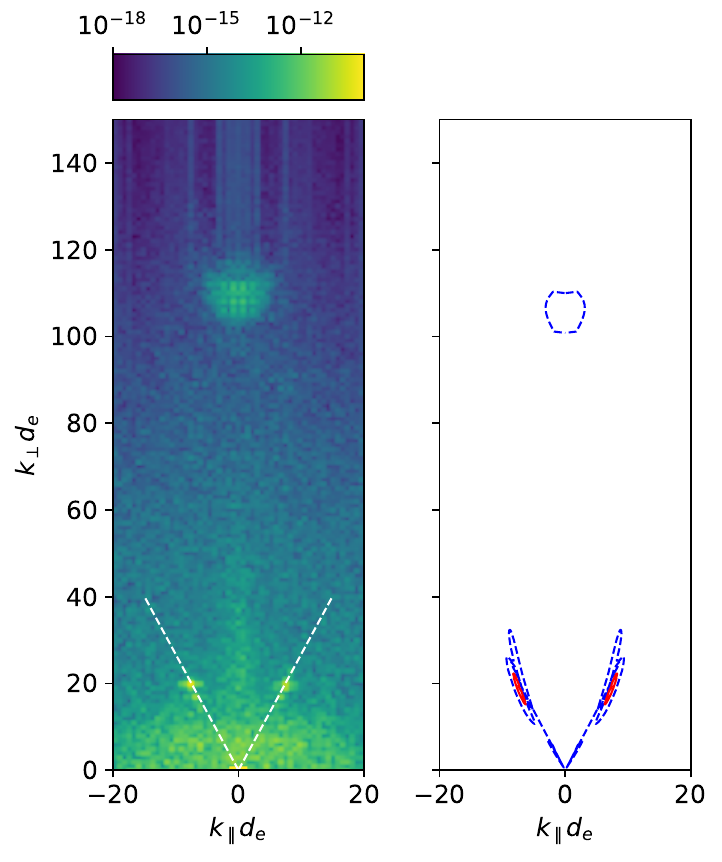}
\caption{Comparison of the wavenumber spectrum of electric field fluctuations measured in the simulation D (left) with the prediction of the linear theory (right).}
\label{fig:2Dspectrum}
\end{figure}

The properties of the electric field fluctuations appearing at $t\Omega_{ce} > 1500$ are consistent with the predictions of the kinetic linear theory introduced above. Figure \ref{fig:2Dspectrum} (left) shows the spectrum of fluctuations of the magnitude of the electric field { $|{\bm E}(\bm k)|^2$} for simulation D, computed at $\Omega_{ce}t \approx 1520$, once the electrostatic fluctuations are beginning to appear. On the right panel, the contour levels for $\gamma/\omega_{pe}=0.001$ (blue) and $\gamma/\omega_{pe}=0.005$ (red) are plotted, obtained from the full dispersion relation \eqref{eq:disp} with $N=3$ sidebands and keeping $\pm 20$ terms in the summation over $n$. The other necessary parameters are extracted from the saturated state of the primary whistler wave before the onset of the secondary instabilities and correspond to $\omega_0=0.35 \Omega_{ce}$ and $V_0/c=2 \times 10^{-3}$. One can see that the linear theory reproduces quite well the range of unstable secondary modes seen in the simulation. The highly oblique whistler waves are excited near the resonance cone indicated by the white dashed lines in Fig. \ref{fig:2Dspectrum} (left), which corresponds to a wave normal angle of $\approx 70^\circ$. Moreover, quasi-perpendicular electron Bernstein are also excited, characterized in Fig. \ref{fig:2Dspectrum} (left) by large values of the perpendicular wave number, $k_\perp d_e \sim 100$. Note that $k_\perp d_e \sim 100$ corresponds to $k_\perp \rho_{eC} \sim 1.25$, i.e. the Bernstein waves have wavelengths of the order of the cold electron gyroradius.

\begin{figure}[h!]
\centering
\includegraphics[width=20pc]{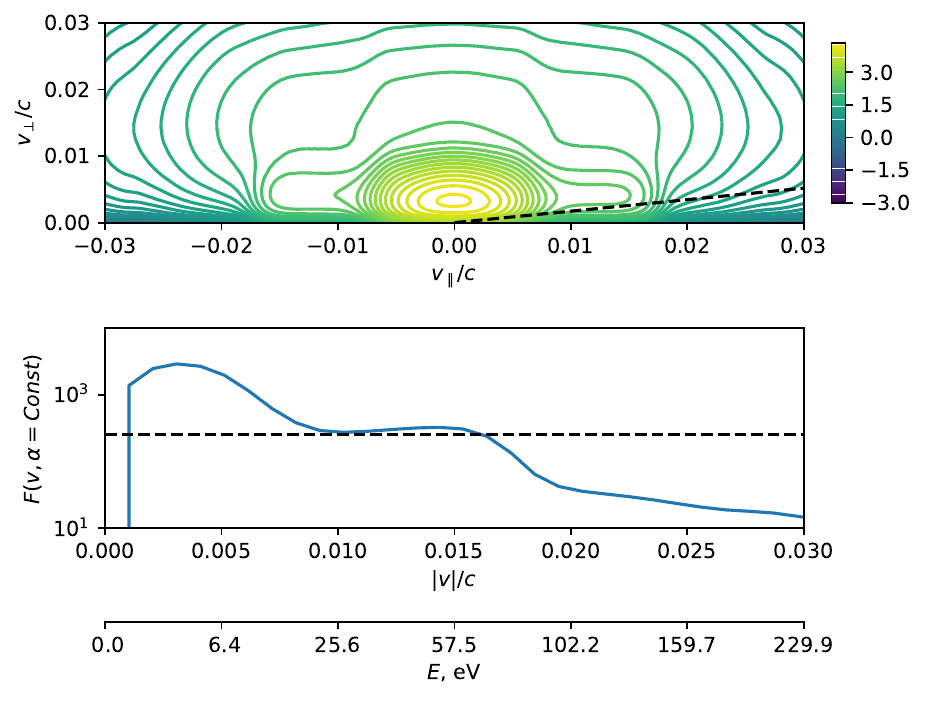}
\caption{Iso-contours of the total electron distribution function $\log F_{e}(v_\parallel,v_\perp)$ (top) and the cut of distribution at pitch-angle $\alpha=10^\circ$ (bottom). The second axis on the bottom panel indicates energy in eV corresponding to the given value of the electron velocity. The horizontal dashed line is included as a visual guide, to highlight existence of a velocity range with a weak positive  gradient in the distribution. }
\label{fig:dist}
\end{figure}

Resonant interaction of the oblique whistler modes with { low-energy} electrons leads to formation of the beam-like features in the {  electron distribution function}. This is illustrated by Figure \ref{fig:dist}, which shows the electron distribution function in the $(v_\perp,v_\parallel)$ space  (top panel) and the distribution function along the cut corresponding to pitch angles $\alpha = 10^\circ$ (bottom panel). The latter roughly corresponds to the angular resolution of the Helium, Oxygen, Proton, and Electron (HOPE) mass spectrometer~\citep{funsten2013} on the Van Allen Probes spacecraft. The energy resolution of the presented diagnostic is however much higher than that of HOPE, enabling unequivocal resolution of sub-100eV structures that may not be picked up by a survey using HOPE. Flattening of the distribution function at parallel energies around 50 eV is apparent.  Using the value of $k_{||}$ corresponding to the peak growth rate from the linear stability analysis, $k_{||} d_e \sim 7$, and the Landau resonance condition with the highly oblique whistler waves at $\omega/\Omega_{ce}\approx 0.35$, $V_{||}=\omega/k_{||}$, leads to $|V_{||}|/c=0.0125$ and a resonant energy of $\sim 40$ eV. This is in good agreement with the position of the plateau. It is also interesting that the distribution function is slightly non-monotonic as a function of parallel energy, reminiscent of a shallow beam. The non-monotonicity is not an artifact of a particular choice of $\alpha$ and is present at all small angles, as well as in the integral of the distribution function over small pitch angles. As was discussed in the introduction, beams are often present concomitant with observations of highly oblique chorus waves. { This led to the hypothesis that the oblique waves are in fact driven by the electron beams}. In the present scenario, the plateau/beam-feature forms due to the secondary drift instabilities and is  the consequence of the appearance of the highly oblique whistler waves.

\begin{figure}[h!]
\centering
\includegraphics[width=20pc]{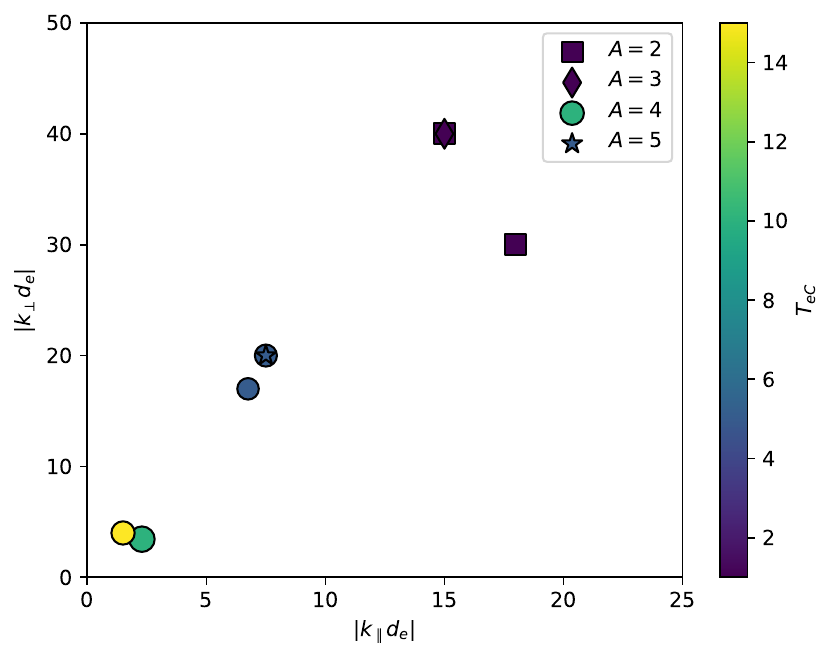}
\caption{Wavenumbers of the spectral peaks corresponding to highly oblique whistler waves observed in the simulations (excluding cases H and I from Table~\ref{tab:params}). The color represents the initial temperature of the cold population $T_{eC}$, while symbols mark cases with the different initial anisotropy of the hot electrons $A$. For a given $T_{eC}$, the amplitude of the excited primary field-aligned waves generally increases with $A$.}
\label{fig:summary}
\end{figure}

The features observed in simulation D are common to all of the simulations in Table \ref{tab:results}. In all cases the highly oblique whistler waves are formed with large wave normal angles near the resonance cone, as shown in Fig. \ref{fig:summary}. The primary whistler waves damp significantly, losing at least $50\%$ of the $|\delta {\bm B}|^2$ acquired after the initial saturation, and in some cases damping almost completely ($\ge 95\%$). This is reported in Table \ref{tab:results}, showing $\Delta B^2=1-{\rm max} |\delta {\bm B}|^2/|\delta {\bm B}(t_{end})|^2$, where ${\rm max} |\delta {\bm B}|^2$ is the maximum of $|\delta{\bm B}|^2=|{\bm B}-{\bm B}_0|^2$ and $t_{end}$ is the final simulation time. Since quasi-linear resonant diffusion rates scale as $|\delta {\bm B}|^2$, such a drastic reduction in wave amplitude also implies a drastic reduction in resonant pitch-angle scattering and energization rates. 

\begin{figure}[h!]
\centering
\includegraphics[width=20pc]{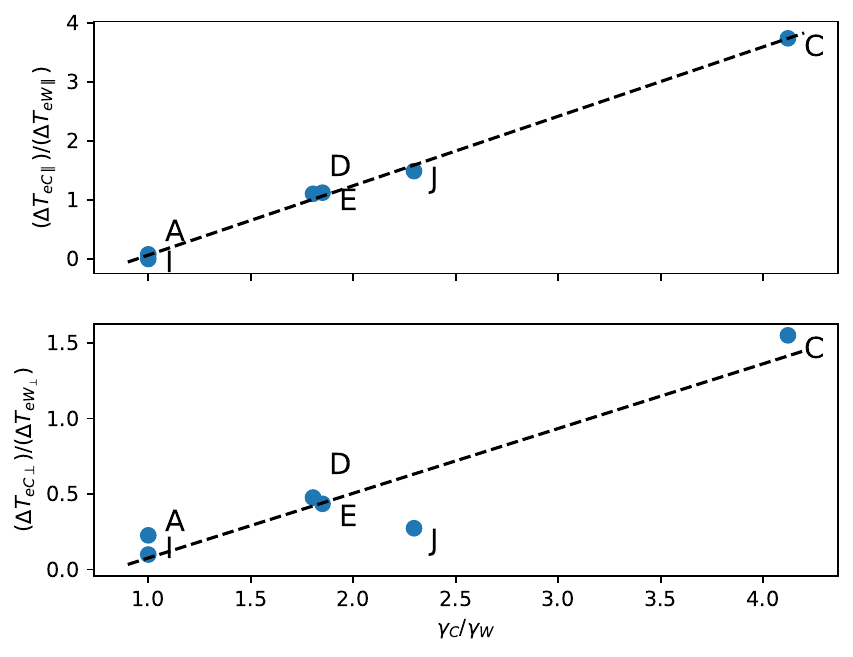}
\caption{Scaling of electron heating with linear growth rate. The figure shows ratios of $\Delta T_{eC}/\Delta T_{eW}$ for parallel (top) and perpendicular (bottom) direction as a function of the ratio between linear growth rate peaks at high and low wavenumbers (see Fig.~\ref{fig:warm_dispersion}). }
\label{fig:dTscaling}
\end{figure}

Part of the primary wave energy is converted into heating of cold/warm electrons, as shown in Table \ref{tab:results} which reports the change in temperature $\Delta T$ in the perpendicular and parallel directions, respectively.
This gives an indication of heating, although we remark that the electron temperature is not always fully saturated at the end of the simulations. An insight into the nature of the energization of background  plasma could be obtained by comparing the ratio of cold-to-warm energy change $\Delta T_{eC}/\Delta T_{eW}$ with the ratio of growth rates at the two peaks of the dispersion relation at small and large values of $k$ discussed in Section~\ref{sec:linear}. This is presented in Fig.~\ref{fig:dTscaling}, where the scaling of $\Delta T_{eC}/\Delta T_{eW}$ with the growth rate ratio is shown for both parallel and perpendicular direction. The growth rates for each case were obtained by numerically solving Eq.~\ref{eq:disp_simple} with the parameters of the corresponding simulations and searching for a local maximum of the growth rate as a function of $k$ and $\theta$ in two ranges of wavenumbers. We excluded cases $B$ and $F$ from the analysis since in each case the numerical heating appears to strongly influence the results. The remaining cases follow a linear scaling for both the parallel and  perpendicular direction.  There is more variability in the perpendicular energization, which may be related to the contribution from the heating by electron Bernstein modes.

We remark that most of the simulations of Table \ref{tab:results} were performed in relatively short spatial domains (admitting only one wavelength of the most unstable modes). The saturation amplitude is generally higher in longer domains, as could be seen by comparing cases A and I. While this limitation may affect specific values of $\Delta B^2$ and $\Delta T$  for given initial conditions, it does not affect the overall conclusions. In fact, one can simply interpret the reported values as being representative of situations where primary waves of a given amplitude have been generated, but the driving mechanism is no longer operational, either because the waves propagated out of the source region, or because the anisotropy of the hot population has reached marginal stability.

\section{Summary and Discussion}

Quasi-field-aligned whistler waves experience a nonlinear instability due to transverse currents when they propagate through a background plasma of sufficiently low energy. One of the most interesting effects associated with the instability is the generation of oblique electrostatic waves, including highly oblique whistler waves near the resonance cone. These modes are of great interest due to their possible connection with oblique whistler-mode chorus waves in the Earth's magnetosphere that are routinely observed by spacecraft and are known to play an important role in wave-particle interactions.  The results presented in this paper extend earlier analysis of~\cite{roytershteyn2021} to the case where low-energy background consists of both a cold electron population with characteristic energy in the range of a few eV and a warm component with characteristic energy in range of $\sim 100$ eV. This is motivated by spacecraft observations in the Earth's magnetosphere where oblique chorus waves were shown to interact resonantly with the warm electrons. 

In the presented analysis the low-energy electrons are modeled by including two (cold+warm) Maxwellian populations. The main conclusions are: i) the instability producing oblique electrostatic waves is sensitive to the shape of the electron distribution at low energies. This is manifested by the sensitivity of both the wavenumbers corresponding to the maximum growth rate and the growth rate magnitude to the temperatures and the relative density of the two low-energy populations. Typically, two distinct peaks in the growth rate are present for the model considered: { a peak at relatively low wavenumbers that is most sensitive to the properties of the warm population and a peak at relatively high wavenumbers that is most sensitive to the properties of the cold population}. The sensitivity of these processes to the low-energy electron distribution function reinforces the need for the development of robust techniques for in-situ measurements in this energy range~\citep{delzanno2021,maldonado2023}; ii) overall, the instability producing oblique whistler waves near the resonance cone persists even in the cases where the   { temperature of the cold populations} is relatively high. For a reference case with (high) amplitude of the parent field-aligned whistler wave $\delta B/B_0 \sim 0.01$,  the maximum growth rate of the oblique whistler mode was $\gamma/\Omega_{ce} \approx 4 \times 10^{-3}$ when the background consisted only of the warm component with temperature equal to 100 eV. iii) PIC simulations show that the development of the instability leads to heating of the { low-energy} electrons and formation of characteristic plateau and beam features in the parallel electron distribution function in the range of energies resonant with the instability. The characteristic energy of the resonant particles changes from 50eV to 1keV for the cases considered depending on the parameters of the background plasma and the driver. The plateau/beam features have been previously detected in spacecraft observations of oblique chorus waves. However, they have been previously attributed to external sources and have been proposed to be one of the mechanisms producing the oblique chorus~\citep{li_2016}. In the present scenario, the causality link is reversed and the oblique whistler waves generated by the instability of the parent quasi-field-aligned whistler are the reason for formation of the plateau and the beam. Finally, (iv) the relative energization of the background populations in the simulations, quantified by the relative change in energy between the cold and the warm populations, have been shown to scale linearly with the ratio of the corresponding peak values of the linear growth rates.  We note that formation of the plateau in the parallel electron distribution functions has  been proposed by~\cite{li_origin_2019} as the mechanism responsible for two-banded structure of the chorus. In that scenario, plateau is formed due to interaction with a slightly oblique primary whistler waves  (in our terminology) and is consequently located at higher energies compared to the case discussed here, where the plateau is formed by interaction with highly oblique secondary whistlers.

In the presented analysis, the effect of drift-type instabilities associated with the presence of cold plasma has been to a large degree isolated by imposing geometrical constraints on the simulations. For example, the simulations were mostly 2D, were performed in relatively small domains with periodic boundary conditions, and considered only uniform background magnetic field and plasma conditions. Relaxing these constraints and changing the configuration towards a more realistic one (e.g. a dipolar magnetic field with non-uniform density) will introduce other important nonlinear phenomena into the problem, such as frequency chirping and the formation of rising or falling chorus elements~\citep{Omura2012}, nonlinear scattering of whistlers~\citep{Ganguli2010},   parametric interaction of whistler waves with electrostatic modes~\citep{Boswell1977,Umeda2017},  and several others, e.g. ~\citep{Artemyev2016,Demekhov2017,Agapitov2018,Vasko2018}. Understanding of the relative significance of the processes discussed in this paper in the presence of other phenomena is an important future direction of this work. 

\section*{Conflict of Interest Statement}

The authors declare that the research was conducted in the absence of any commercial or financial relationships that could be construed as a potential conflict of interest.



\section*{Funding}

GLD was supported by the Laboratory Directed
Research and Development (LDRD) - Directed Research Program and Exploratory Research Program
of Los Alamos National Laboratory (LANL) under project number 20200073DR and 20220453ER.
JH was supported by LANL through its Center for Space and Earth Science (CSES). CSES is funded by the LDRD program under project number 20210528CR.
LANL is operated by Triad National Security,
LLC, for the National Nuclear Security Administration of
U.S. Department of Energy (Contract No. 89233218CNA000001).
VR was partially supported by NSF award 2031024 and NASA grant 80NSSC22K1014.

\section*{Acknowledgments}
The authors gratefully acknowledge discussions with George Hospodarsky.

This research used resources of the National Energy Research Scientific Computing Center (NERSC), a U.S. Department of Energy Office of Science User Facility located at Lawrence Berkeley National Laboratory, operated under Contract No. DE-AC02-05CH11231, resources provided by the NASA High-End Computing (HEC) Program through the NASA Advanced Supercomputing (NAS) Division at Ames Research Center, and resources at the Texas Advanced Computing Center (TACC) at The University of Texas at Austin. 

\section*{Data Availability Statement}
The raw data supporting the conclusions of this article will be made available by the authors, without undue reservation. 

\bibliographystyle{Frontiers-Harvard} 
\bibliography{whistler}


\section*{Tables}

\begin{table*}[h!]
\caption{Parameters of the PIC simulations. $\omega_{pe}/\Omega_{ce}=4$ and $n_i/n_0=1$ in all cases.}
\label{tab:params}
\begin{tabular}{cccccccccccc}
\hline
Sim & ${L}$ & $N$ & $\Delta t$ & $n_{eC}$ & $T_{eC}$ & $T_{eW}$ &$n_{eW}$ & $n_{eH}$ & $T_{eH||}$  & $A^H$ & $T_i$   \\
    &    [$d_e$]      &     & [$\omega_{pe}^{-1}$] & [$n_0$] & [eV] & [eV] & [$n_0$] & [$n_0$] & [keV] & & [eV]\\
\hline
A & $5.48 \times 5.48$ & $768 \times 768 $ &  $4.5 \times 10^{-3} $ & 0.5 & 10 & 100 &0.3 & 0.2   & $2$ & 4 & 10   \\
B & $0.63 \times 8.38$ & $384 \times 4864$  & $1.2 \times 10^{-3}$  & 0.75 & 1  &  100  & 0.2 & 0.05 & $10$ & 2 & 1   \\
C & $0.63 \times 8.38$ & $384 \times 4864$  & $1.2 \times 10^{-3}$  & 0.75 & 1  &  100  & 0.2 & 0.05 & $10$ & 3 & 1   \\
D & $6.28 \times 8.38$ & $1600 \times 2160$ & $2.7 \times 10^{-3}$  & 0.75 & 5  &  100  & 0.2 & 0.05 & $10$ & 4 & 5   \\
E & $6.28 \times 8.38$ & $1600 \times 2160$ & $2.7 \times 10^{-3}$  & 0.75 & 5  &  100  & 0.2 & 0.05 & $10$ & 5 & 5  \\
F & $6.28 \times 8.38$ & $960 \times 1260$  & $ 4.6 \times 10^{-3}$ & 0.75 & 15 &  100  & 0.2 & 0.05 & $10$ & 4 & 15 \\
G & $0.63 \times 6.3$ & $540 \times 5000$   & $8.4 \times 10^{-4}$  & 0.8  & 1  & - & 0     & 0.2  & $2$  & 2 & 1   \\
H  & $ 0.63^2 \times 5.48$ & $48^2 \times 512 $ & $6.8 \times 10^{-3}$  & 0.8& 10 & - & 0   & 0.2  & $2$  & 4 & 10  \\
I  & $ 5.48 \times 10.96$ & $768 \times 1536$ & $5 \times 10^{-3}$  & 0.5 & 10  &  100  & 0.3 & 0.2   & $2$ & 4 & 10  \\
J & $6.28 \times 8.38$ & $1600 \times 2160$ & $2.7 \times 10^{-3}$  & 0.75 & 5  &  150  & 0.2 & 0.05 & $10$ & 4 & 5   \\
\hline
\end{tabular}
\end{table*}

\begin{table*}[h!]
\caption{Results of PIC simulations. Notes: 1) heating in cases B and F is strongly affected by numerical errors; 2) simulation H is 3D.
}
\label{tab:results}
\begin{tabular}{cccccccc}
\hline
Sim &  $\Delta B^2$  & $\Delta T_{eC||}$ &  $\Delta T_{eC\perp}$  &  $\Delta T_{eW||}$ &  $\Delta T_{eW\perp}$  & $\theta_*$ & $k_*$ \\
&   [\%] &  [eV] &  [eV] &  [eV] &  [eV] & [deg] & [$d_e$] \\
\hline
A &  99  & 0   & 3.7    & 72.3 & 37.4 & 56 & 4\\
B &  65  & 0.1 & 0      & 0  & 0.1  & 69 & 43 \\
C &  95  & 1   & 0.2   & 0.25  & 0.1  & 69 & 43 \\
D &  95  & 2.4  & 0.3   & 2.1  & 0.1  & 69 & 21 \\
E &  97  & 2.8   & 0.3   & 2.5  & 0.8  & 69 & 21\\
F &  70  & 0.2  & 0.1    & 1 & -0.2 & 69 & 43\\
G &  82  & 0.3  & 0.3   & -  & -  & 59 &  35 \\
H &  95  & 25  &  20 & -  & - \\
I &  100  & 10.7 & 15.9  & 136.8 & 70.4 & 58 & 5\\
J &  95   & 2.2 & 0.3 & 1.5 & 1 &  68 & 18 \\
\hline
\end{tabular}
\end{table*}

\section*{Figure captions}


\end{document}